%% file: sample-sigconf.tex
\documentclass[sigconf,screen]{acmart}

\usepackage{amsfonts}       
\usepackage{nicefrac}       
\usepackage{paralist}
\graphicspath{ {./figures/} }

\AtBeginDocument{%
  }

\setcopyright{acmlicensed}
\copyrightyear{2026}
\acmYear{2026}
\acmDOI{XXXXXXX.XXXXXXX}
\acmConference[Graphics Interface '26 (Under Review)]{Make sure to enter the correct
  conference title from your rights confirmation email}{June 09--12,
  2026}{Kitchener-Waterloo, Ontario}
\acmISBN{978-1-4503-XXXX-X/2026/06}

\begin{document}

\title[SpringTime]{SpringTime: Learning Simulatable Models of Cloth with Spatially-varying Constitutive Properties}

\author{Guanxiong Chen}
\email{gxchen@cs.ubc.ca}
\affiliation{%
  \institution{University of British Columbia}
  \city{Vancouver}
  \country{Canada}
}

\author{Shashwat Suri}
\affiliation{%
  \institution{University of British Columbia}
  \city{Vancouver}
  \country{Canada}
}

\author{Yuhao Wu}
\affiliation{%
  \institution{University of British Columbia}
  \city{Vancouver}
  \country{Canada}
}

\author{Yixian Cheng}
\affiliation{%
  \institution{University of British Columbia}
  \city{Vancouver}
  \country{Canada}
}

\author{Ganidhu Abeysirigoonawardena}
\affiliation{%
  \institution{University of British Columbia}
  \city{Vancouver}
  \country{Canada}
}

\author{Etienne Vouga}
\email{evouga@cs.utexas.edu}
\affiliation{%
  \institution{University of Texas at Austin}
  \city{Austin}
  \country{USA}
}

\author{David I.W. Levin}
\email{diwl.levin@utoronto.ca}
\affiliation{%
  \institution{University of Toronto}
  \city{Toronto}
  \country{Canada}
}

\author{Dinesh K. Pai}
\email{pai@cs.ubc.ca}
\affiliation{%
  \institution{University of British Columbia}
  \city{Vancouver}
  \country{Canada}
}

\renewcommand{\shortauthors}{Chen et al.}

\begin{abstract}
  Materials used in real clothing exhibit remarkable complexity and spatial variation due to common processes such as stitching, hemming, dyeing, printing, padding, and bonding. Simulating these materials, for instance using finite element methods, is often computationally demanding and slow. Worse, such methods can suffer from numerical artifacts called ``membrane locking'' that makes cloth appear artificially stiff. 
  Here we propose a general framework, called SpringTime, for learning a simple yet efficient surrogate model that captures the effects of these complex materials using only motion observations. The cloth is discretized into a mass-spring network with unknown material parameters that are learned directly from the motion data, using a novel force-and-impulse loss function. Our approach demonstrates the ability to accurately model spatially varying material properties from a variety of data sources, and immunity to membrane locking which plagues FEM-based simulations. Compared to graph-based networks and neural ODE-based architectures, our method achieves significantly faster training times, higher reconstruction accuracy, and improved generalization to novel dynamic scenarios. Codebase for the paper can be found at \url{https://github.com/ericchen321/springtime}.
\end{abstract}


\keywords{Do, Not, Use, This, Code, Put, the, Correct, Terms, for,
  Your, Paper}

\begin{teaserfigure}
  \includegraphics[width=\textwidth]{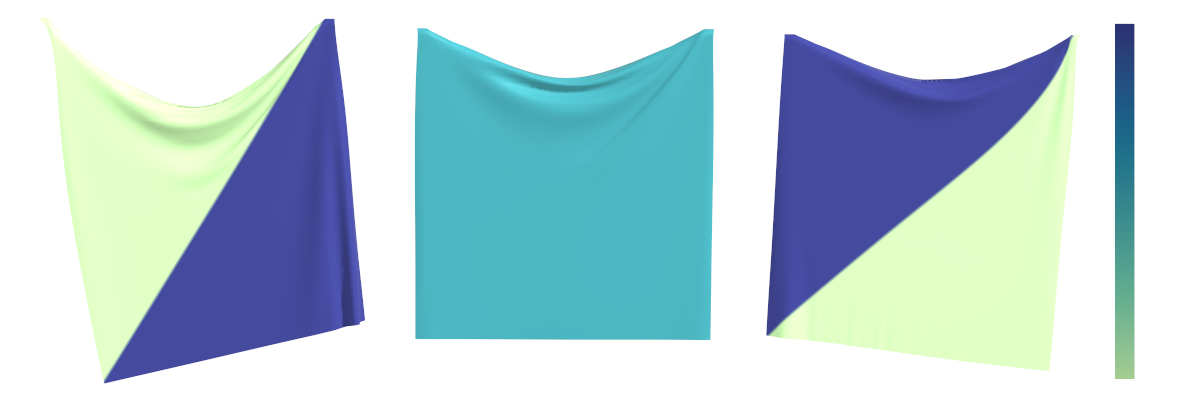}
  \caption{Equilibrium configurations of a square piece of cloth with spatially heterogeneous (left, right) vs. using homogeneous material (middle). Purple regions are stiffer than yellow ones. For the homogeneous cloth all triangles share the same stiffness, as in previous work, which is taken to be the average stiffness of triangles in the heterogeneous cloth. Despite identical initial and boundary conditions, stiffness variation leads to distinct behaviors. Our method can capture such variation.}
  \label{fig:hetero_matters}
\end{teaserfigure}

\received{20 February 2007}
\received[revised]{12 March 2009}
\received[accepted]{5 June 2009}

\maketitle

\section{Introduction}
\input{sections/intro}

\section{Related Work}
\input{sections/related_work}

\section{Method}
\input{sections/method}

\section{Experiments}
\input{sections/experiments}

\section{Conclusion}
\input{sections/concl}
\clearpage
\bibliographystyle{ACM-Reference-Format}
\bibliography{sample-base}

\appendix
\section{Synthetic Motion Clip Set Construction}
\label{sec:clipset_con}
\input{sections/app_syn_data_gen}

\section{Implementation Details and Hyperparameters}
\label{sec:impl_details}
\input{sections/app_impl}

\section{Additional Results}
\label{sec:add_results}
\input{sections/app_add_results}

\end{document}

%% file: sections/intro.tex
Simulation of physical systems is a cornerstone of modern science and engineering, as well as physically based animation in computer graphics, games, and VR. The Finite Element Method (FEM) is widely used in computer graphics, to predict the behavior of everything from cars to characters. However, modeling and simulation of real-world objects remains challenging for several reasons. We will focus on the cloth simulation in this paper, but the issues are more general.

First, real objects may have complex, spatially varying, constitutive properties that significantly affect behavior. A simple example is shown in Fig.~\ref{fig:hetero_matters}.  A striking example is the Issey Miyake A-POC clothing design (generated from ``A Piece of Cloth'') demonstrated at SIGGRAPH Asia 2024\footnote{https://blog.siggraph.org/2024/11/siggraph-asia-2024-keynote-yoshiyuki-miyamae.html/}. Heat-shrinking different parts of cloth creates spatially varying constitutive properties that gives the cloth the desired 3D shape. Even common processes such as stitching, printing, and bonding introduce spatially varying material properties that are challenging to model using standard methods and by previous data-driven techniques. 

Second, FEM cloth simulations can be complex to implement and slow. As a consequence, FEM is rarely used in real-time applications, and it is common to resort to using coarse meshes and low order (linear) elements. 

\begin{figure*}[ht]
    \centering
    \includegraphics[width=\textwidth]{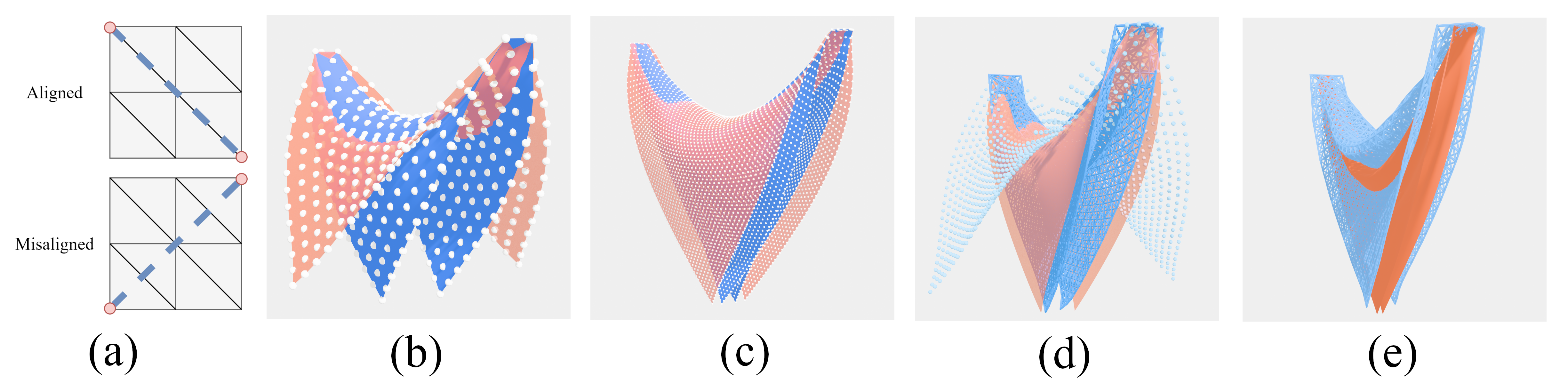}
    \caption{The folding experiment. A square piece of cloth is pinned at diagonally opposite corners, in two ways.  (a) Illustration with 2x2 mesh, with pinned corners (red) and the fold line (dashed blue). In the ``aligned'' case, the fold is aligned with mesh edges; membrane locking occurs when it is ``misaligned.'' (b) Aligned (blue) vs misaligned (red) low-resolution ($16 \times 16$) mesh at an equilibrium configuration. The misaligned case appears stiffer due to locking. (c) Aligned (blue) vs misaligned (red) high-resolution ($61 \times 61$) mesh at an equilibrium configuration. The differences due to locking are reduced. (d) SpringTime (blue mass-spring sheet) vs low-resolution mesh landmarks used for training  (blue dots) vs high-resolution mesh (red) in misaligned rest configuration. Even though SpringTime was trained on the low-resolution mesh, its behavior is closer the high resolution mesh, with less locking. (e) SpringTime trained on mesh with bending stiffness (blue) vs high-resolution mesh without bending stiffness (red), in the aligned configuration. This shows that SpringTime can learn real bending stiffness while resisting locking. White dots are mesh vertices; all meshes have the same mass and physical dimensions, and only differ in bending stiffness or resolution.}
    \label{fig:folding}
\end{figure*}

Third, and most insidiously, FEM meshes typically used in computer graphics can suffer greatly from numerical artifacts such as {\em membrane locking}, which lead to unrealistic bending behavior. Fig.~\ref{fig:folding} (b) and (c) illustrates this phenomenon. 
In engineering, locking is ameliorated by using higher-order elements, reduced integration, or with special elements designed specially for shells, which add significantly to the complexity of FEM simulation.


In this work, we introduce a learning-based framework, called \textit{SpringTime}, 
that addresses all these problems. For run-time simulation, SpringTime uses a simple mass-spring network cloth model that is easy to implement and efficient to simulate in real-time applications. Moreover, we show that SpringTime can effectively learn spatially-varying constitutive properties of cloth from simulated experiments using a more complex and slow model (e.g., an FEM model). Thus SpringTime may be used as a real-time ``surrogate'' model for a more complex off-line model. 

Even though there is a popular perception that mass-spring networks are unsophisticated and less appropriate for modeling continuum mechanics, that is not necessarily the case. As observed by \citet{breen1994predicting}, Hearle and colleagues at Manchester's renowned Department of Textiles had investigated applications of shell and plate models in the 1980s, and concluded: ``But, in dealing with the three-dimensional buckling of textile fabrics, neither the terminology nor the methodology of the established mechanical theory of the bending of plates and shells is of much help because of the limiting assumptions that are made'' \citep{amirbayat1989anatomy}. Mass-spring networks could better capture the anisotropic behavior of cloth, and avoid numerical artifacts of FEM discretization of thin shells. We demonstrate, with a series of experiments, that SpringTime is resistant to the membrane locking problem that plagues FEM simulations, while avoiding the complexity of special solutions. Remarkably, we observe that SpringTime behaves better under bending than even the FEM model that it was trained with, thereby achieving superresolution in inference. We illustrate these phenomena and offer some potential explanations.

\begin{figure}[h]
    \centering
    \includegraphics[width=.49\textwidth]{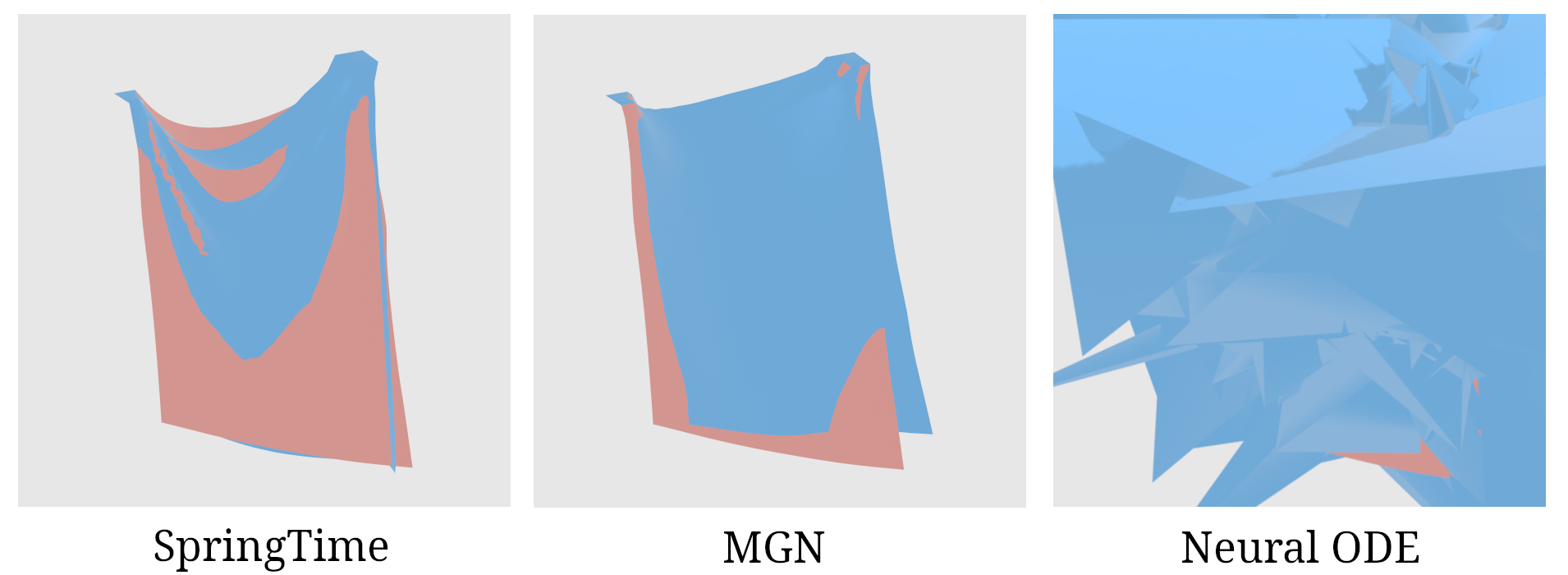}
    \caption{Results from different surrogate models (blue) after $8$ seconds of simulation, compared with the ground truth (red). (Left) Our surrogate model;  (middle), (right) are popular alternatives. Notice that SpringTime yields comparable steady-state reconstruction error as MeshGraphNet (MGN, middle), while Neural ODE (right) completely fails to reconstruct long rollouts.}
    \label{fig:gen_comp_with_bls}
\end{figure}

The main contribution of this paper is the mass-spring-based framework consisting of a modern stochastic learning pipeline with a physically based mass-spring (rather than neural) network. Specific features include:
\begin{itemize}
    \item A novel force-and-impulse-based loss function;
    \item The ability to learn heterogeneous models with spatially-varying constitutive properties;
    \item Resistance to membrane locking problems that plague FEM-based simulations;
    \item Minimal requirements - it needs only point cloud data and total mass of the cloth for training, ground-truth cloth mesh not required;
    \item Flexibility in defining the resolution of the  surrogate model.
\end{itemize}
The framework produces results that match or exceed state-of-the-art neural approaches in terms of accuracy and generalizability (see Fig.~\ref{fig:gen_comp_with_bls}), while requiring significantly less training time.


%% file: sections/related_work.tex

\textbf{Neural constitutive models.} A continuing challenge with deformable object simulation is defining an approriate consitutivie  (material) model. Off-the-shelf models and parameters can be applied, but for more complicated objects, optimizing the constitutive model to agree with real-world measurements or specialized numerical data is common~\citep{pai2001scanning,bickel2009capture,miguel2012data, wang2011data}. More recently, deep learning has been deployed for this task. Graph Neural Networks~\citep{pfaff2020learning, sanchez2020} or U-Nets~\citep{Bhaduri2022} replace the entire simulation stack, spatial discretization and time integration, with learned update rules. These methods have been applied to cloth simulation, rigid body simulation with contact and granular material simulation, showing impressive results replicating scenarios similar to their training data. Rather than replacing both the spatial and temporal components of a simulation, Neural ODEs~\citep{chen2018neural} choose to retain a classical, numerical time integrator while learning the integrand of the system (in our case the coupled first order dynamics). Neural constitutive models which learn offsets~\citep{Wang2020, chen2022crom, chang2023licrom} or entire neural replacements~\citep{Ma2023} to existing continuum mechanics-based material models have been applied to volumetric objects (not thin shells as we do here), while Neural Jacobian~\citep{aigerman2022neural} fields can learn static deformations (not dynamic like our work). Rather than making any neural substitutions, our work retains the stochastic training machinery of deep learning approaches but applies it to a physical model, in our case a mass-spring network. 

Mass-spring systems are often regarded as outdated, but they offer important advantages over their more widely used discrete counterparts. Most notably, mass-spring networks can represent volumetric, thin-shell, and rod-like structures within a unified discretization, depending on the network topology and the assigned per-spring material parameters~\citep{da2015new, kot2015elastic, liu2013fast, xu2022efficient, golec2020hybrid}. Despite their efficiency and expressiveness, deriving parameter values from classical constitutive models or estimating these parameters from data~\citep{da2015new, bianchi2004simultaneous, Bianchi2003} has been historically challenging, especially due to the complexity of constitutive properties in heterogeneous materials~\citep{Ju2024, zhong2024springgaus}.  Closest to our method is Spring-Gaus~\citep{zhong2024springgaus}, which fits a mass-spring-like network to volumetric Gaussian Splats. However, the employed node-based storage of material parameters allows for the same spring to produce forces that are not  equal and opposite on its end-points, thus violating Newton's third law. 

\textbf{Cloth Simulation and constitutive property estimation.} Simulations pertaining to cloth have always been a prime subfield in graphics and simulation research due to their varied industrial applications, from yarn-level simulators by \citet{Sperl2022,Cirio2014} which specialize in woven cloths, to continuum-level or meshed-based simulators~\citep{shao2025learning, Ju2024, grigorev2023hood, Kairanda2023, zhao2023learning, Feng2022, li2022diffcloth, Santesteban2022, Bertiche2022, Li2022, pfaff2020learning, Liang2019, wang2011data}. 
Many of these works, including~\citet{Ju2024, Feng2022, wang2011data} focus on accurate constitutive parameter estimates by \textit{real-to-sim} adaptations using specialized measurement devices; others~\citep{shao2025learning, grigorev2023hood, pfaff2020learning} focused more on learning constitutive properties from synthetic data, and demonstrated great generalizability to real-world data~\citep{zheng2024physavatar, li2022diffcloth, Li2022, Liang2019}, or more specifically, learning garment deformation models jointly with human motion kinematics~\citep{stuyck2023diffxpbd, zhao2023learning, grigorev2023hood, li2024diffavatar, Bertiche2022, Santesteban2022}. However, cloth or garments chosen as the subject to estimate haven been almost exclusively homogeneous, composed of a single fabric without complex fine-scale patterns and structures; the only exceptions are ~\citet{stuyck2023diffxpbd} and ~\citet{grigorev2023hood}, yet both explores fitting specifically of garments on virtual try-on's. The focus of our work is to address the broader problem of estimating any piece of cloth's \textit{heterogeneous} constitutive properties, such as stiffness that varies across space using synthetic data. We distinguish from work like~\citep{stuyck2023diffxpbd, Li2022, pfaff2020learning} that our estimation pipeline does not require a reference mesh as input, and from ~\citep{zheng2024physavatar, grigorev2023hood, stuyck2023diffxpbd, Bertiche2022} that we focus on the broader domain of modelling constitutive properties of a standalone cloth, rather than the specific domain of high-quality reconstruction of garment kinematics on moving human avatars. We show that by fitting a physically correct mass-spring network using our novel loss and training curriculum, we achieve comparable or better accuracy and generalizability than previous approaches but at a significant reduction in training time and model complexity. 

\textbf{Membrane locking.} For simulating dynamic elastica, the general consensus is that finite element spatial discretizations~\citep{sifakis2012fem} coupled with implicit time integration~\citep{li2020incremental} are state-of-the-art, with specialized discretization schemes available for thin sheets~\citep{narain2014arcsim} or rods~\citep{bergou2008discrete}. Finite element methods are rooted in continuum mechanics~\citep{mase2009continuum} and so allow for a wide-range of mathematically or experimentally derived material models which have been extensively validated across many domains. Yet a continuing challenge pertaining to FEM-based simulations is membrane locking~\citep{stolarski1982membrane}: when a finite-element mesh is not finely discretized, under certain configurations the mesh would appear to experience larger bending resistance due to some phantom stiffness. Previous work on FEM-based cloth simulation have attempted to address this issue by either simulating up to fine scales~\citep{zhang2022progressive, narain2012adaptive} or applying constraints to solving dynamics~\citep{chen2019locking, jin2017inequality, bender2011simulating}. While adaptive remeshing techniques used in recent garment simulators~\citep{grigorev2023hood, pfaff2020learning} can address membrane locking to some extent, they incur higher computational cost in training and inference; Mass-spring models which, under appropriate stiffness and resolution definition are less prone to suffer from membrane locking, on the other hand, can be a natural choice for combating the problem. This motivates us to build a mass-spring based system, and we show that our trained surrogate does not exhibit membrane locking that FEM-based simulations suffer from.

%% file: sections/method.tex
\label{sec:method}
\begin{figure*}[ht]
    \centering
    \includegraphics[width=\textwidth]{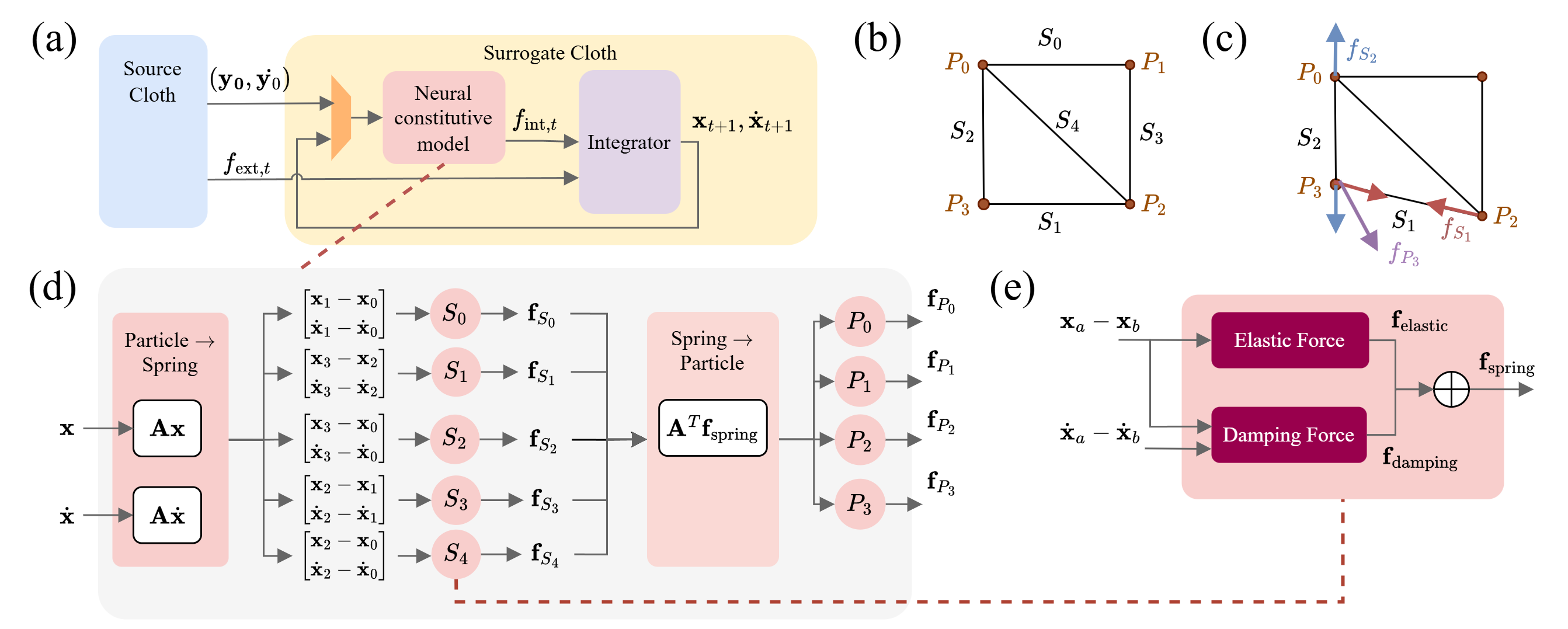}
    \caption{Method overview. (a) The material modeling pipeline. We sample system state $(\mathbf{y}_0, \dot{\mathbf{y}_0})$ from the source cloth at the first time step of a rollout, resample following the scheme in Sec.~\ref{sec:spatial_disc} to get target particle positions, feed it to the neural constitutive model $h_{\theta}$ to predict neural particle force $\mathbf{f}_t$, then integrate. To further evolve the surrogate we use previously predicted states. (b)-(e) illustrates the architecture of our SpringTime which models a simple rectangular cloth, with rest configuration shown in (b) and a deformed configuration in (c). (d): Each neural spring (circle) takes the relative position and velocity of particles at its two ends, and predicts the spring's internal force. Then each particle accumulates forces applied by its surrounding springs. (e) Neural spring. At every time step, the module computes elastic and damping force and sums them up. Parts with learnable parameters are in dark red.}
    \label{fig:method_overview}
\end{figure*}

\textbf{Learning task definition.} 
Fig.~\ref{fig:method_overview} shows the neural surrogate modeling pipeline and the architecture of our neural constitutive model. Our method takes as input a \emph{source system}: $T$ frames of motion of a piece of cloth, with $N$ landmark points on the cloth tracked from frame to frame. In addition to the positions $\mathbf{y}_i \in \mathbb{R}^{3N}$ of the landmarks at each frame $i$, we assume the source system also provides the external force $\mathbf{f}_{\text{land},i}$ acting on each landmark at each frame (e.g. gravity), an estimate of the cloth's area density $\rho$, and a binary classification of each landmark as either free or pinned in place. This source system data might come from a high-fidelity finite element simulation or from video observations of a real-world sample. We train a surrogate model to capture the behavior and constitutive properties of the source system while obeying the laws of physics. Once trained on the $T$ input frames, the model can be used to efficiently and accurately predict the motion of the cloth given novel initial conditions, external forces, and boundary conditions. The surrogate system is a mass-spring network consisting of a user-specified number of particles $P$ connected by $S$ springs. The task is to learn constitutive parameters (stiffness and damping) of each spring from the source system's landmark trajectories. As a preprocessing step, we map the given landmark trajectories to target positions $\hat{\mathbf{x}}_i \in \mathbb{R}^{3P}$ of each particle, and the external forces to forces $\mathbf{f}_{\text{ext},i}$ on each particle (Section~\ref{sec:spatial_disc}). These quantities are used to supervise training of the neural constitutive model at the heart of our surrogate system, which predicts the nonlinear restoring and damping forces $\mathbf{f}_{\text{springs}}$ exerted by the springs given the surrogate system's current position and velocity (Section~\ref{sec:neural_const_model}):
\begin{equation}
    h_{\theta}\left(\begin{bmatrix}
        \mathbf{x}^T\\
        \dot{\mathbf{x}}^T
        \end{bmatrix}\right): \mathbb{R}^{6P} \rightarrow \mathbb{R}^{3S}.
\end{equation}
We map the predicted spring forces to internal forces on each surrogate particle~\cite{liu2013fast},
\begin{equation}
    \mathbf{f}_{\text{int}} = \left(\mathbf{A} \otimes I_{3\times 3}\right)^T\mathbf{f}_{\text{springs}},
\end{equation}
where $\mathbf{A} \in \mathbb{R}^{S\times P}$ is the sparse signed incidence matrix of the mass-spring network (with $A_{sa}=-1$, $A_{sb}=1$ if spring $s$ connects particle $a$ to particle $b$).

Finally, given these internal forces and the prescribed external forces and boundary conditions, we use semi-implicit Euler time integration to advance the state of the surrogate model from frame to frame. We learn the neural constitutive model parameters $\theta$ by comparing this simulated trajectory to the ground-truth motion observations $\mathbf{\hat{x}}_{i=1:T}$ (Section~\ref{sec:train_curriculum}).


\subsection{Spatial Discretization and Resampling}
\label{sec:spatial_disc}

\begin{figure}[ht]
    \centering
    \includegraphics[width=.4\textwidth]{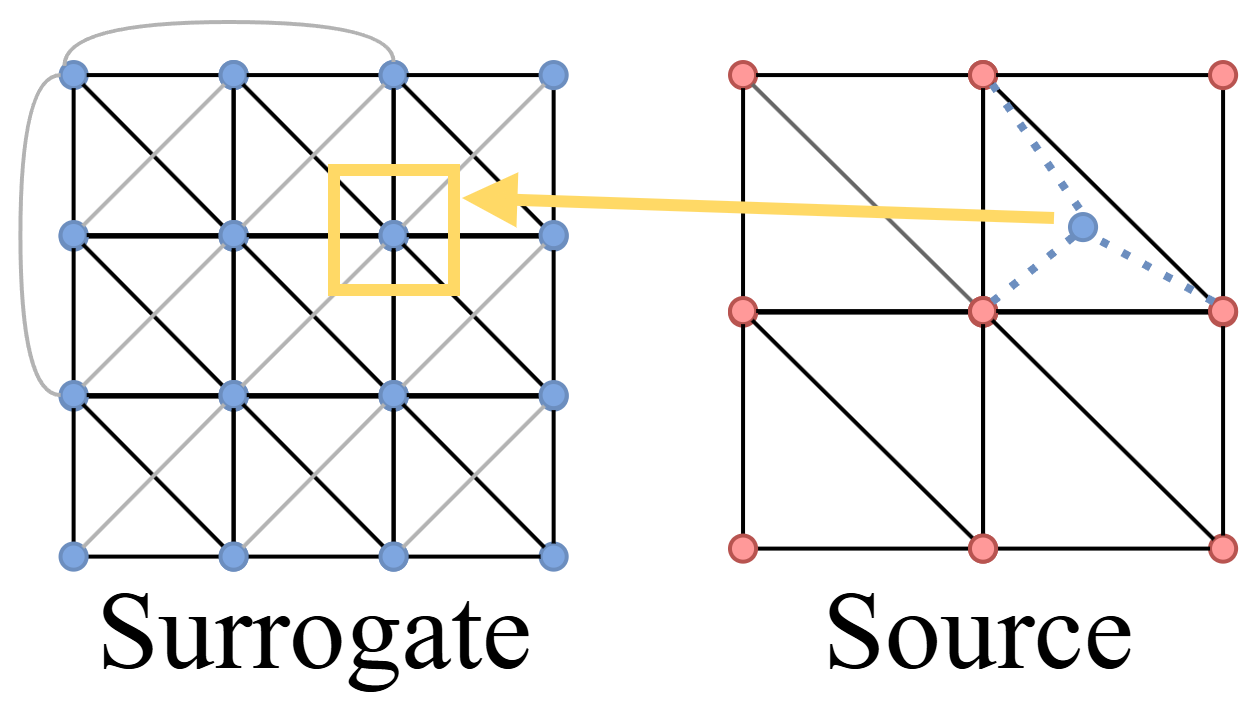}
    \caption{Spatial discretization of the surrogate system (left) and the source system (right). Bending springs and diagonal springs from top-left to bottom-right can be optionally excluded on construction. For simplicity, we only show two bending springs; the full surrogate contains bending springs connected at $\text{stride}=1$. The dotted lines show how we compute the target position of each surrogate particle from the positions of nearby source landmarks via barycentric interpolation. Note that in training only landmarks from the source are needed, and topolgical information are not necessary.}
    \label{fig:spatial_disc}
\end{figure}

Given a desired number of surrogate particles $P$, we build our surrogate mass-spring network as follows: we isometrically unroll the source system's cloth specimen into a rectangle in the plane, then discretize this rectangle with a $P$-vertex regular square grid, as shown in Fig.~\ref{fig:spatial_disc} (left). We place a surrogate spring at each edge of the grid, with the spring rest length $l_0$ determined by the edge length in this rest configuration.

As shown in Fig.~\ref{fig:spatial_disc} (right), the source system landmarks do not necessarily correspond to the surrogate particles (and the resolution of the surrogate particle grid might be much coarser or finer than the density of landmark points). Therefore we must resample the given landmark positions $\mathbf{y}_i$ at each frame $i$ to positions of the surrogate particles.

Let $\bar{\mathbf{y}}$ and $\bar{\mathbf{x}}$ be the rest positions of the unrolled landmarks and surrogate particles within the 2D rectangle, respectively. For each surrogate particle $p$, we identify its three nearest neighbors $v_{\{1,2,3\}}$ among the $\bar{\mathbf{y}}_j$. Let $b_{\{1,2,3\}}$ be the barycentric coordinates of $\bar{\mathbf{x}}_p$ within triangle $\left\{\bar{\mathbf{y}}_{v_1}, \bar{\mathbf{y}}_{v_2}, \bar{\mathbf{y}}_{v_3} \right\}$. We set the target position $\mathbf{\hat{x}}_{p,i}$ of particle p at frame $i$ to
\begin{equation}
    \mathbf{\hat{x}}_{p,i} = b_0 \mathbf{y}_{v_0,i} + b_1 \mathbf{y}_{v_1,i} + b_2 \mathbf{y}_{v_2,i}.
\end{equation}
We map external forces to the surrogate particles using barycentric interpolation in an analogous way.

We also need to assign a mass $m_p$ to each surrogate particle. We simply set $m_p = \rho A / P$, where $A$ is the rectangle area.

\subsection{Neural Constitutive Modeling}
\label{sec:neural_const_model}


As shown in Fig.~\ref{fig:method_overview} (d), we use the incidence matrix $\mathbf{A} \otimes I_{3\times 3}$ to extract the displacement $\mathbf{d}$ and relative velocity $\mathbf{v}$ between each spring's two endpoints from the system position and velocity vectors $\mathbf{x}$ and $\dot{\mathbf{x}}$, as described in~\citet{liu2013fast}'s method of fast simulation of mass-spring systems. The total internal force applied by each neural spring is computed as the sum of an elastic restoring force and viscous damping force (see Fig.~\ref{fig:method_overview} (e)):
\begin{equation}
    \label{eq:total_neural_force}
    \mathbf{f}_{\text{spring}} = \frac{k (\|\mathbf{d}\| - l_0)}{\|\mathbf{d}\|}\mathbf{d} + \frac{b (\mathbf{v} \cdot \mathbf{d})}{\|\mathbf{d}\|^2}\mathbf{d} ,
\end{equation}
where the elastic and damping coefficient $k_{s=1:S}$, $b_{s=1:S}$ are learnable parameters.

\subsection{Simulation and Training}
\label{sec:train_curriculum}

\textbf{Forward dynamics.} We set the surrogate system's initial conditions based on the target positions computed in Section~\ref{sec:spatial_disc}, i.e. 
$$\mathbf{x}_0 = \hat{\mathbf{x}}_0, \quad \dot{\mathbf{x}}_0 = \frac{\hat{\mathbf{x}}_1-\hat{\mathbf{x}}_0}{\Delta t},$$
where $\Delta t$ is the source system time step. We step these initial conditions forward in time with semi-implicit Euler integration. In particular, we update velocity using
\begin{equation}
    \mathbf{M} (\dot{\mathbf{x}}_{j+1} - \dot{\mathbf{x}}_{j}) = \Delta t \,\left[\mathbf{f}_{\text{ext},j} + \mathbf{f}_{\text{springs}}(\mathbf{x}_j)\right],
\end{equation}
where $\mathbf{M}$ is the $\mathbb{R}^{3P \times 3P}$ diagonal mass matrix, and position using $\mathbf{x}_{j+1} = \mathbf{x}_{j} + \Delta t\, \dot{\mathbf{x}}_{j+1}$.

\textbf{Loss formulation.} 
We iterate the above time integration process to simulate an entire trajectory $\mathbf{x}_{j=1:T}$. We compare this trajectory to the target trajectory $\hat{\mathbf{x}}_{j=1:T}$ using the loss
\begin{equation}
    \label{eq:total_loss}
    L = \lambda_{\text{f}} L_{\text{f}} + \lambda_{\text{J}} L_{\text{J}} + \lambda_{\text{k neg.}} L_{\text{k neg.}} + \lambda_{\text{b neg.}} L_{\text{b neg.}}.
\end{equation}
In Eq.~\ref{eq:total_loss}, $L_{\text{f}}$ denotes the force loss, which penalizes difference between target and simulated net force applied on each particle in the system at each time step. The target net force is estimated from the target positions and Newton's Second Law:
\begin{align}
    L_{\text{f}} &= \frac{1}{P(T-2)}\sum_{i=2}^{T-1} ||\mathbf{f}_{i} - \mathbf{\hat{f}}_{i} ||_2^2\\
    \mathbf{f}_{i} &= \mathbf{M} \left(\frac{\mathbf{x}_{i+1} - 2 \mathbf{x}_i + \mathbf{x}_{i+1}}{\left(\Delta t\right)^2}\right),
\end{align}
where $\mathbf{f}_{i} = \mathbf{f}_{\text{ext},i} + \mathbf{f}_{\text{springs}}(\mathbf{x}_i)$ denotes the net force (including gravity, environmental damping, and the spring forces) applied to each particle at time step $i$ of forward dynamics. $L_{\text{J}}$ denotes the impulse loss, which penalizes difference between target and predicted system impulse:
\begin{equation}
    L_{\text{J}} = \frac{1}{P} \left\|\mathbf{J} - \mathbf{\hat{J}}\right\|_2^2,
\end{equation}
where $\mathbf{J}$ is the accumulated impulse on each particle from the start to the end of the trajectory, integrated using the Trapezoid Rule:
\begin{equation}
    \mathbf{J} = \frac{1}{2} \sum_{j=1}^{T-1} (\mathbf{f}_{j} + \mathbf{f}_{j+1}) \Delta t,
\end{equation}
and $\hat{\mathbf{J}}$ is computed analogously from the $\hat{\mathbf{f}}$.
Intuitively, the force and impulse loss terms act analogously to the $P$ and $I$ terms in a PID controller, where the former penalizes discrepancy between predicted and target instantaneous forces applied to each particle and the latter penalizes error accumulation over time. We notice through experiments that the optimal weights $\lambda_{J}$, $\lambda_{f}$ vary across different sources of data (i.e. source systems) and need to be fine-tuned, just as the weights of a PID controller must be tuned when a mechanical system's parameters change.

The final two loss terms of Equation~\eqref{eq:total_loss} are regularization terms that penalizes non-physical, negative spring stiffness or damping constants. Specifically,
\begin{align}
    L_{\text{k neg.}} &= \sum_{s=1}^{S}\operatorname{ReLU}(-k_s)\\
    L_{\text{b neg.}} &= \sum_{s=1}^{S}\operatorname{ReLU}(-b_s),
\end{align}
where the $k$ and $b$ are the learned per-spring parameters of our neural constitutive model (see Sec.~\ref{sec:neural_const_model}). The term stays at zero as long as the estimated parameter stays positive. 

Many alternative loss functions for training the surrogate could also be used, such as position loss~\citep{Ma2023} or acceleration loss~\citep{sanchez2020}, but in our ablation experiments (Sec.~\ref{sec:exp_ablation}) we found that the force-and-impulse loss works best in our setting.

\textbf{Training from multiple trajectories.} For clarity of exposition, we've assumed above that the source system consists of only one trajectory with $T$ frames of motion. Our approach extends in a straightforward way to training from multiple source system trajectory samples, where the same surrogate mass-spring network is used to fit all trajectories.

\textbf{Dual-pass training.} While the elastic force is only position-dependent, damping force depends on both particle position and velocity. To disentangle the effects of each force on particle dynamics, we devise a training curriculum that isolates learning stiffness from learning damping by splitting the training into two passes. In the first pass we freeze damping weights and optimize only the stiffness weights, and in the second pass we freeze stiffness weights and optimize only damping weights. Furthermore, we use low-velocity motion data sampled from the source system in the first pass of training, so non-gravitational force applied on particles will be dominated by the elastic force. We discuss more on how low-velocity data is sampled in Sec.~\ref{sec:clipset_con}.

%% file: sections/experiments.tex
\subsection{Synthetic Data Generation}
\label{sec:exp_syn_data_gen}

\textbf{Constructing source systems.} We build two source systems of cloths modeled under different simulation frameworks:
\begin{inparaenum}[(i)]
    \item A square-shaped cloth made up of particles and springs. Forces that act upon particles come from springs only;
    \item A square-shaped cloth constructed from a triangular mesh. Internal forces are evaluated based on FEM principles. We use NVIDIA WARP's FEM-based cloth simulation publicly available on Github~\citep{warp2022}.
\end{inparaenum}
We discretize each cloth into a $P \times 3P$ particle grid. The choice of $P$ varies across experiments; for the experiment in which we compare SpringTime with baselines on generalizability to novel boundary and initial conditions, we use $P=31$, which is close to the discretization scale picked by recent work such as~\citet{shao2025learning}. We assign different stiffness values to either springs in the mass-spring cloth or triangles in the mesh-based cloth so they have spatially-varying constitutive properties: see Fig.~\ref{fig:hetero_matters}. 

\begin{figure}[ht]
    \centering
    \includegraphics[width=.49\textwidth]{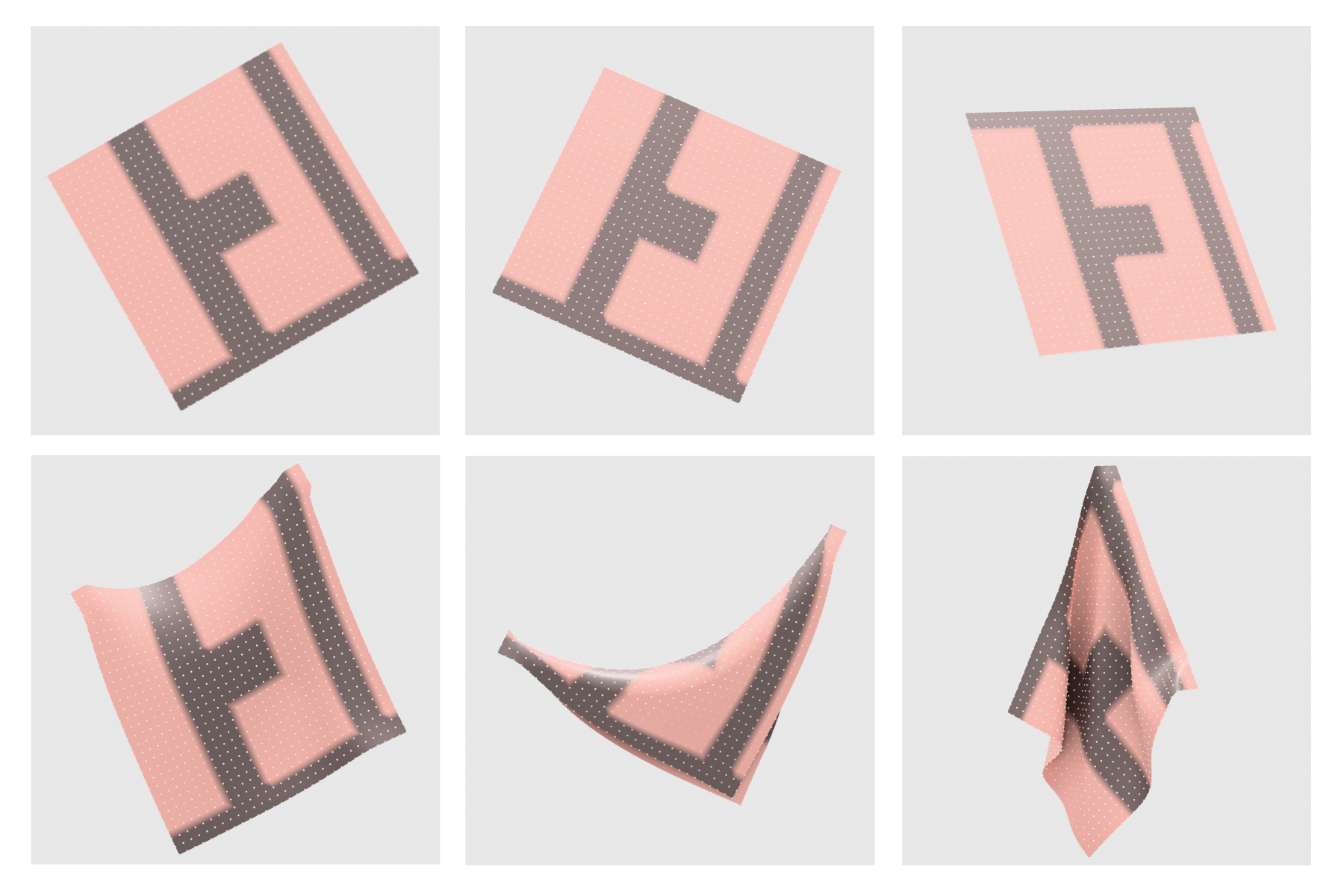}
    \caption{Generating $N=3$ rollouts of equal lengths but with different initial conditions (top) and boundary conditions. Equilibrium states reached by the last step (bottom). Stiffer regions are colored darker.}
    \label{fig:syn_data_gen}
\end{figure}

\textbf{Generating rollouts with different initial and boundary conditions}. To collect ground-truth simulation data, for each cloth we anchor a set of its vertices so they cannot move, and we simulate the deformation of the rest of the cloth over time. The set of vertices being fixed in space, formally defined as anchor neighborhood $\Omega$ defines the boundary condition of a rollout; the initial position and velocity of all particles $\{\mathbf{y}, \mathbf{\dot{y}} \}$ defines its initial condition. For each rollout $n$, we randomly pick two anchor neighborhoods as boundary condition $\Omega_n$, and we apply a random rotation to the sheet to assign the cloth with initial condition $\{ \mathbf{y}_n, \dot{\mathbf{y}}_n \}$: see Fig.~\ref{fig:syn_data_gen}. For training SpringTime and the Neural ODE baseline we use $N=512$ rollouts; for testing we use $N=16$ rollouts. Each rollout lasts $8$ seconds, during which the cloth falls under the influence of gravity and Rayleigh damping force $\mathbf{f}_{\text{Rayleigh}} = -b_{\text{Rayleigh}}\dot{\mathbf{x}}$, where Rayleigh damping constant $b_{\text{Rayleigh}}=0.1$. We simulate with timestep size of $1~\mathrm{ms}$, and this time discretization scheme gives us $8~\text{k}$ time steps of data from each rollout.

\textbf{Constructing motion clip dataset.} We follow the approach in prior work to break down rollouts into shorter clips~\citep{Ma2023, gartner2022differentiable}, then train on a batch of clips in each iterations. We refer readers to Sec.~\ref{sec:clipset_con} for more details.

\subsection{Metrics}

In some of our experiments, we evaluate SpringTime by applying it to a source system that is itself a simulation of a mass-spring network. In this setting, we can easily compare the surrogate's estimated spring stiffness and damping parameters to those of the source system. We quantify the error in the reconstructed parameters by computing the root-mean-squared error (RMSE) between their estimated and ground-truth values:
\begin{equation}
    \text{RMSE}_{k} = \sqrt{\frac{1}{S} \sum_{s=1}^{S} \left(k_s - \hat{k}_s\right)^2}.
    \label{eq:rmse_k}
\end{equation}
where $\hat{k}_s$ is the ground-truth stiffness constant in the source system. For the damping constants we compute $\text{RMSE}_{b}$ analogously. These two metrics offer us an objective measure of how well the surrogate model captures the source system's constitutive properties.


When the source system is a finite element simulation or a video of real-world cloth, ground-truth spring parameters are not available. We instead assess the quality of our surrogate model by comparing its simulated motion to the source system trajectory. Specifically, we compute the RMSE between the estimated and target positions of each particle at trajectory frame $j$:
\begin{equation}
    \text{Motion RMSE}_{j} = \sqrt{\frac{1}{P} \sum_{i=1}^{P} \left(\mathbf{x}_{j,i} - \hat{\mathbf{x}}_{j,i}\right)^2},
    \label{eq:rmse_motion}
\end{equation}
where $\mathbf{x}_{j,i}$ denotes the position of the $i$-th particle at frame $j$ and $\hat{\mathbf{x}}_{j,i}$ is the corresponding target position (computed via interpolation of the landmarks of the source system as described in Sec.~\ref{sec:spatial_disc}). For an entire trajectory or ensemble of trajectories, we compute the mean and standard deviation of the Motion RMSE across all time steps of all trajectories.

\subsection{Reconstructing Spatially Varying Materials}
The goal of this experiment is to show that SpringTime can accurately estimate the stiffness and damping properties of a source cloth with spatially varying stiffness. We use the $\text{RMSE}_{k}$ and $\text{RMSE}_{b}$ metric established in Eq.~\ref{eq:rmse_k} to assess the quality of our constitutive property estimates, and to use them, we build the source cloth with masses and springs, and we construct a surrogate system with the same resolution as the source cloth. Bending and the complete set of shear springs are included in both the source and the surrogate. This allows us to establish a one-to-one mapping from springs in the source cloth to springs in the surrogate cloth. We also assess the quality of motion reconstruction using the $\text{Motion RMSE}$ metric from Eq.~\ref{eq:rmse_motion}. 

\begin{figure}[ht]
    \centering
    \includegraphics[width=.49\textwidth]{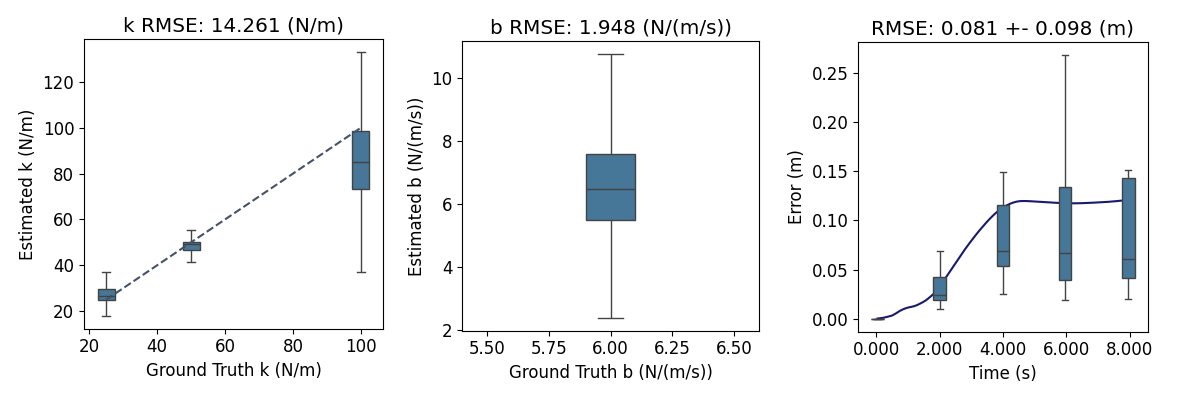}
    \caption{Predicted vs ground-truth (left) spring stiffness constants; (middle) damping constants; (right) per-frame motion reconstruction RMSE. The curve shows mean RMSE across $16$ test rollouts on each time step; we subsample five steps to visualize the distribution across rollouts. Each cloth contains $5,826$ springs.}
    \label{fig:recon_acc}
\end{figure}

Fig.~\ref{fig:recon_acc} shows the accuracy of stiffness and damping estimates, and the reconstruction error obtained from a test set which contains $16$ $8~\mathrm{s}$-long rollouts. The source cloth we used to generate training and test rollouts is a square-shaped, $15.5  \times 15.5~\mathrm{m}$ cloth discretized into a $32 \times 32$ particle grid connected by springs. The source cloth contains springs with three stiffnesses: $25$, $50$ and $100 ~\mathrm{N/m}$, and all springs share the same damping constant of $6 \ \mathrm{N/m/s}$. Fig.~\ref{fig:recon_acc} (left) tells that most of our estimated stiffnesses are close to the ground-truth. The damping estimates are also close to the ground-truth on average, despite some variations: see Fig.~\ref{fig:recon_acc} (middle).

\subsection{Robustness against Membrane Locking}
\label{sec:memb_lock}
\begin{figure}[ht]
    \centering
    \includegraphics[width=.49\textwidth]{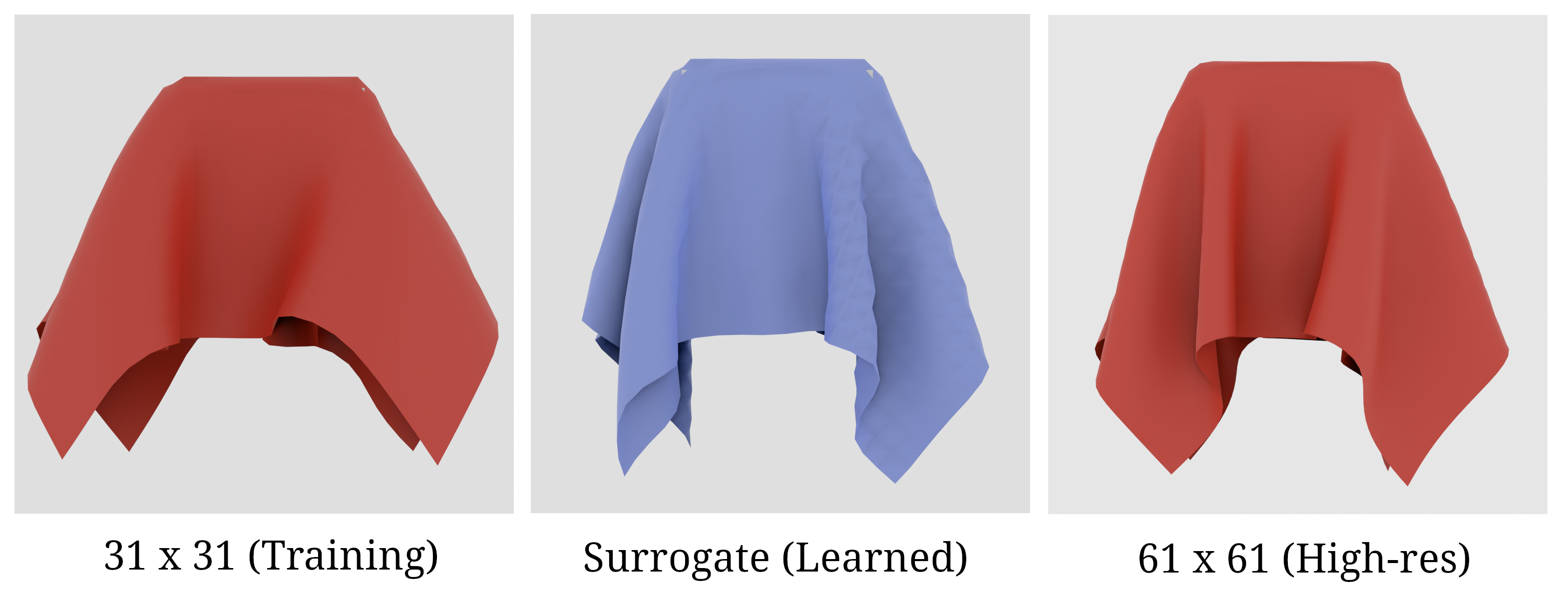}
    \caption{Our 32x32 surrogate (center) was trained on a data from a 31x31 FEM mesh (left) that exhibits significant membrane locking. This is apparent in the drape test of the cloth on a square table. Remarkably, our results more closely match the results of a higher resolution FEM cloth simulation (right), indicating that our model captures material properties well.}
    \label{fig:train_vs_high_res}
\end{figure}

Finite element methods are susceptible to membrane locking, a phenomenon that produces artificial bending stiffness that is more significant at coarse resolutions~\citep{stolarski1982membrane}. An interesting consequence of our approach, which is learning material properties from force and impulse losses and not the shape of the draped cloth, is that we are able to model material properties even when the training data generated from a low-resolution FEM-based simulation is significantly affected by the numerical limitations of membrane locking. Thus, our simulation results match higher-resolution FEM simulations more closely than the training data-- see Fig.~\ref{fig:train_vs_high_res}.

To further validate the super-resolution effect of the mass-spring-based surrogate, as shown in~Fig.~\ref{fig:folding} (b), (c), we pin two corners of a square-shaped FEM-simulated mesh along a diagonal, make the cloth fold under the influence of gravity and Rayleigh damping, and sample the cloth's equilibrium configurations. We compare the equilibrium configurations under two cases:
\begin{inparaenum}[(i)]
    \item ``Aligned'', i.e. folding along the major diagonal that coincides with diagonal edges in the mesh;
    \item ``Misaligned'', i.e. folding along the minor diagonal.
\end{inparaenum}
Under the misaligned case, as we expect, the cloth sags less than the aligned case, and the difference in equilibrium configurations is more apparent on the low-resolution mesh. As shown in Fig.~\ref{fig:folding} (d), our mass-spring-based surrogate which includes the full set of diagonal and bending springs, and trained on landmark trajectories sampled from the low-resolution mesh in (b),  however is robust against this effect, because we can see that at equilibrium configuration it is closer to the high-resolution mesh that suffers less from membrane locking. Further, by comparing Fig.~\ref{fig:folding} (d), (e) we can see that the surrogate is robust against the pseudo-stiffness caused by membrane locking not by chance, but because it can properly fit to the actual bending resistance of the mesh--- the surrogate at equilibrium configuration is closer to the mesh with bending stiffness (d) than the mesh with zero bending stiffness (e). 

\begin{figure}[ht]
    \centering
    \includegraphics[width=.49\textwidth]{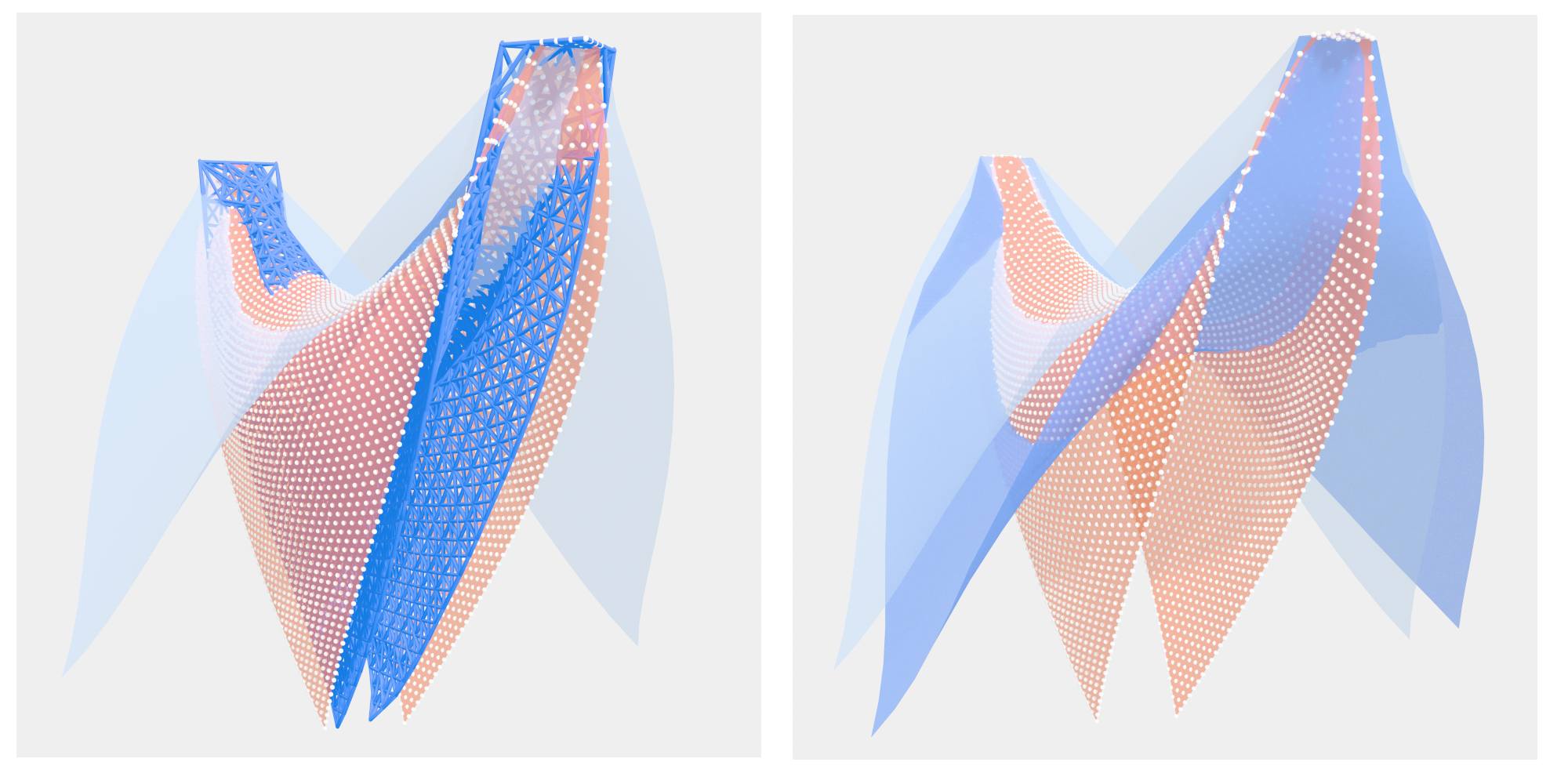}
    \caption{SpringTime (left, dark blue) vs MGN (right, dark blue) trained on kinematic data sampled from low-resolution ($16 \times 16$) mesh (light blue), settled to the misaligned configuration in evaluation. A high-resolution ($61 \times 61$) mesh (red) is placed in each scene for reference.}
    \label{fig:springtime_vs_mgn}
\end{figure}

Moreover, we offer a qualitative comparison with MGN~\citep{pfaff2020learning} in terms of the two surrogate's behavior in the misaligned equilibrium configuration. Notice that in Fig.~\ref{fig:springtime_vs_mgn} MGN matches the low-resolution mesh's behavior more than the high-resolution mesh's, which makes sense, because the surrogate is trained to model the behavior of the low-resolution mesh when boundary conditions and trajectories are provided as inputs.

While both SpringTime and MGN are data-driven methods that encode cloth's constitutive properties and reconstruct reference motion in inference, what makes SpringTime stands out is its robustness to numerical artifacts that exist in the training data. We believe SpringTime attains the robustness by more accurately estimating the cloth's constitutive properties, after seeing trajectories sampled in a multitude ($512$) rollouts, many of which would not lead to a ``locked'' equilibrium configuration as in the misaligned case. MGN on the other hand is more vulnerable to such artifacts in the training data, since it tends to memorize the locked equilibrium configuration corresponding to the input boundary condition. We refer readers to Sec.~\ref{sec:app_hparams} for hyperparameters used to train MGN.

\subsection{Generalization Experiments}
Here we discuss two experiments that assess how well SpringTime generalizes to novel scenarios not included in training rollouts. Moreover, we demonstrate that despite our neural constitutive model is built from masses and springs, it can be trained to emulate the behavior of source cloths constructed from triangular meshes bearing a different resolution and simulated with FEM principles.

\begin{table}
\caption{Comparison of our method with prior differentiable cloth simulators with publicly available codebases across key features. $\times$ indicates a method not supporting the feature natively; $\circ$ indicates native support provided but without demonstration; \checkmark indicates demonstrated native support. Res-independence means support for simulating surrogate at finer or more coarse resolution than reference data; Long rollout means simulating rollouts longer than 2 seconds.}
\label{tab:all_baselines}
\scriptsize
\centering
\begin{tabular}{ccccc}
\textbf{Method} & \textbf{Heterogeneity?} & \textbf{Res-indep.?} & \textbf{Ref-mesh free?} & \textbf{Long rollouts?} \\
\hline
\textbf{Ours} & \checkmark & \checkmark & \checkmark & \checkmark \\
\textbf{NCLaw~\citep{Ma2023}} & $\circ$ & $\times$ & $\times$ & $\circ$ \\
\textbf{DiffCloth~\citep{li2022diffcloth}} & $\times$ & $\times$ & $\times$ & \checkmark \\
\textbf{MGN~\citep{pfaff2020learning}} & $\circ$ & $\circ$ & \checkmark & \checkmark \\
\textbf{GNN~\citep{sanchez2020}} & $\circ$ & $\times$ & \checkmark & $\circ$ \\
\textbf{\citet{Liang2019}} & $\times$ & $\times$ & $\times$ & \checkmark \\
\textbf{Neural ODE~\citep{chen2018neural}} & $\circ$ & \checkmark & \checkmark & $\circ$ \\
\end{tabular}
\end{table}

\textbf{Baselines.} Tab.~\ref{tab:all_baselines} compares our method with several recent, open-sourced differentiable cloth or general deformable simulators across key features--- if the simulator supports modelling heterogeneous cloths, can simulate surrogate bearing different spatial resolution from the source, if it needs a reference mesh to kick-start training and if it supports simulating long rollouts. We focus exclusively on methods that allow us to model material properties of a stand-alone piece of cloth from kinematic data and optionally, the total mass and physical dimension of the system. While state-of-the-art garment simulators as in~\citep{shao2025learning, li2024diffavatar, grigorev2023hood, Santesteban2022} has achieved stunning results in reconstructing garment kinematics on moving human avatars, we exclude them as they tackle the specific body-and-garment joint fitting problem. In Sec.~\ref{sec:memb_lock} we have made a brief qualitative comparison with~\citet{pfaff2020learning} in terms of robustness to membrane locking in simulation data. Here we compare, both quantitatively and qualitatively between our surrogate system and the following baseslines, in terms of ability to generalize to new initial and boundary conditions:
\begin{inparaenum}[(i)]
    \item Neural ODE~\citep{chen2018neural} which is an obvious choice for modelling constitutive properties, due to the simplicity and flexibility of its MLP-based architecture --- see Fig.~\ref{fig:arch_ours_vs_mlp} in Sec.~\ref{sec:add_results} for an illustration of the architecture;
    \item MeshGraphNet (MGN)~\citep{pfaff2020learning} which can learn a surrogate model for complex thin-shell materials from point cloud data.
\end{inparaenum}
We exclude state-of-the-art neural constitutive methods like NCLaw~\citep{Ma2023} from being compared as baselines, since they require premium knowledge such as internal stress in their training curricula to learn complex 3D deformables. Another criterion for our comparisons is the method's ability to model heterogeneity in cloth, so we have excluded methods like~\citet{Liang2019, li2022diffcloth} which places greater emphasis on modelling homogeneous cloth.
We demonstrate that our neural constitutive model is able to generalize better to scenes with novel initial or boundary conditions, and it adapts well to scenes that involve interaction of new objects over extended period of time. We also compare SpringTime with the two selected baselines in terms of training time and model complexity, measured in number of learnable parameters.

\begin{table}
    \caption{Motion reconstruction RMSE of our method vs the baseline across test rollouts of different lengths. We report mean and standard deviation of per-frame reconstruction RMSE across $200$ frames, $16$ rollouts for $8~\mathrm{s}$ long simulations.}
    \label{tab:gen_novel_ic_bc_results}
    \begin{center}
    \begin{tabular}{lccc}
    \textbf{Time} & \textbf{SpringTime} & \textbf{MGN} & \textbf{Neural ODE} \\
    \hline
    $8~\mathrm{s}$  & $\mathbf{0.64 \pm 0.42}$ & $0.68 \pm 0.45$ & $6.34 \pm 6.30$ \\
    \end{tabular}
    \end{center}
\end{table}

\begin{figure}[ht]
    \centering
    \includegraphics[width=.49\textwidth]{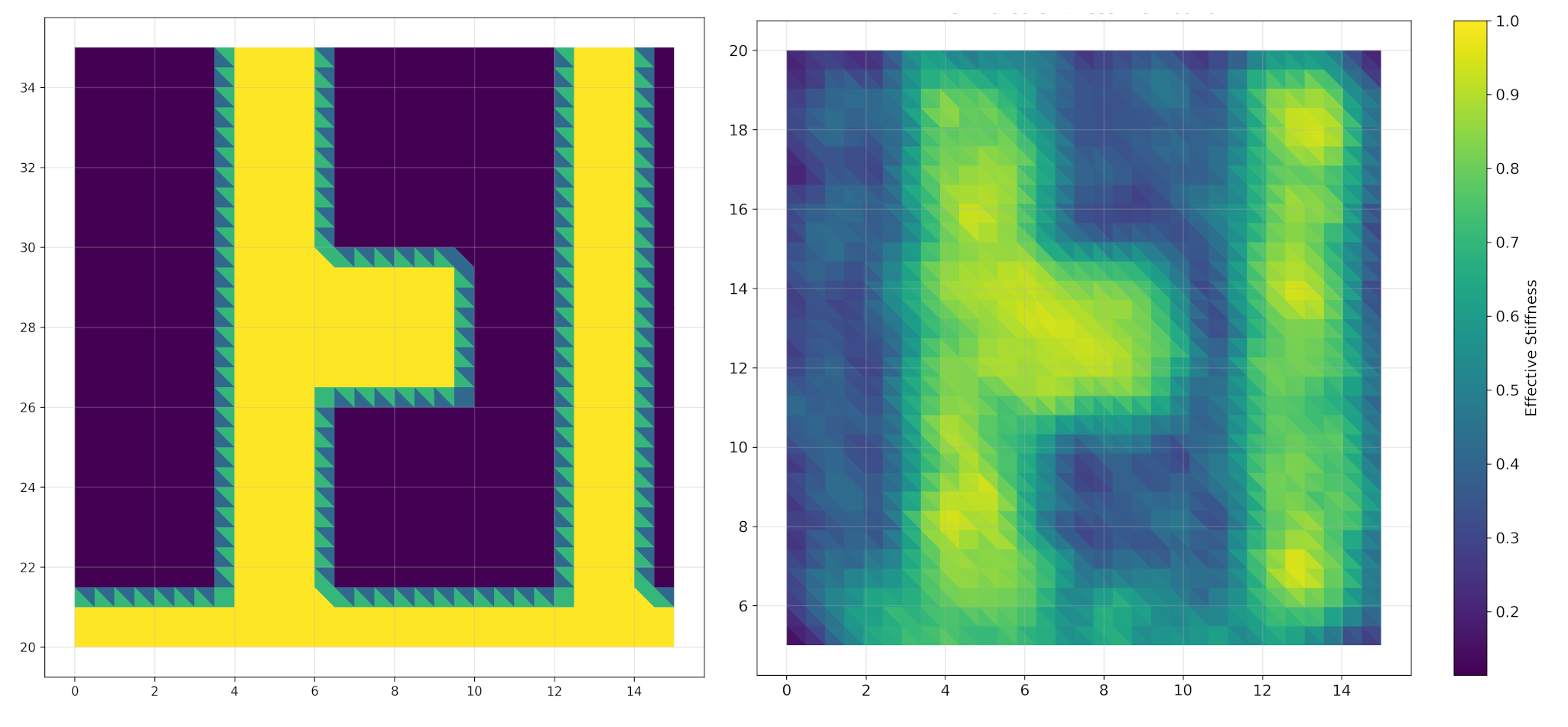}
    \caption{Distribution of material stiffness for the (left) The mesh-based source cloth; (right) learned mass-spring-based surrogate cloth. The stiffness of the two cloth are normalized to the same scale from $0$ to $1$. Darker regions are more stiff.}
    \label{fig:stiffness_distribution}
\end{figure}

\textbf{Generalization to novel initial and boundary conditions.} We train SpringTime and the baselines on $512$ rollouts generated from a triangular mesh-based cloth with spatially-varying stiffness: see Fig.~\ref{fig:stiffness_distribution} (left). The mesh-based cloth is discretized into a $31 \times 31$ particle grid. For running SpringTime and Neural ODE, we discretize the surrogate cloth into a $32 \times 32$ particle grid, which is relatively prime to the resolution of the source cloth. MeshGraphNet on the other hand requires the source cloth's default topological structure for training and testing, so we do not perform resampling. We then train SpringTime and the baselines, and use the trained models to reconstruct $16$ test-set rollouts with new initial and boundary conditions, visualize the reconstructed rollouts and compute reconstruction error reported in Tab.~\ref{tab:gen_novel_ic_bc_results}. We see SpringTime outperforms both baselines in terms of motion reconstruction accuracy over long rollouts.
Fig.~\ref{fig:stiffness_distribution} (right) shows the spatial variation in stiffness in the fitted surrogate cloth. Notice that despite the source cloth is built from meshes rather than springs and masses, SpringTime is still able to recover the two stiffer stripes that run from the top to the bottom and the central patch, although reconstruction of the bottom, horizontal stripe is sub-optimal.
In terms of qualitative assessment, we show in Fig.~\ref{fig:gen_comp_with_bls} that after simulating for $8 \mathrm{s}$ SpringTime can attain a steady state closer to the ground-truth than the baselines. We refer readers to Fig.~\ref{fig:gen_novel_ic_bc} for more qualitative results. 
We see from the motion RMSE in Tab.~\ref{tab:gen_novel_ic_bc_results} and Fig.~\ref{fig:gen_comp_with_bls} that Neural ODE surrogate completely disintegrates after $8 \mathrm{s}$ of simulation. We hypothesize this behavior to be due to lack of topological information available to the model. As the surrogate evolves over time, lacking of topological information allows the particles to roam freely, thereby disintegrating the cloth. Methods that use MLPs to learn dynamics~\citep{Ma2023}, do so along with other strong priors as mentioned in the previous section.

Turning to training time and model complexity: without leveraging domain-specific information, both baselines require more learnable parameters to learn the topology of a cloth: while SpringTime contains $11,652$ learnable parameters to fit a surrogate system with resolution of $32 \times 32$, Neural ODE requires over $6\text{M}$ learnable parameters, and MGN contains $2.3\text{M}$ parameters. We also measured the training time of SpringTime and the baselines on the same platform, and found that training SpringTime for two passes requires only $120$ minutes, while Neural ODE requires $100$ minutes of training time which is slightly less than SpringTime. MGN on the other hand requires considerably more training time: training for $3.8M$ iterations requires $48$ hours, and we expect training for $10M$ steps as used in~\citep{pfaff2020learning} would require $126$ hours.

\begin{figure}[ht]
    \centering
    \includegraphics[width=.49\textwidth]{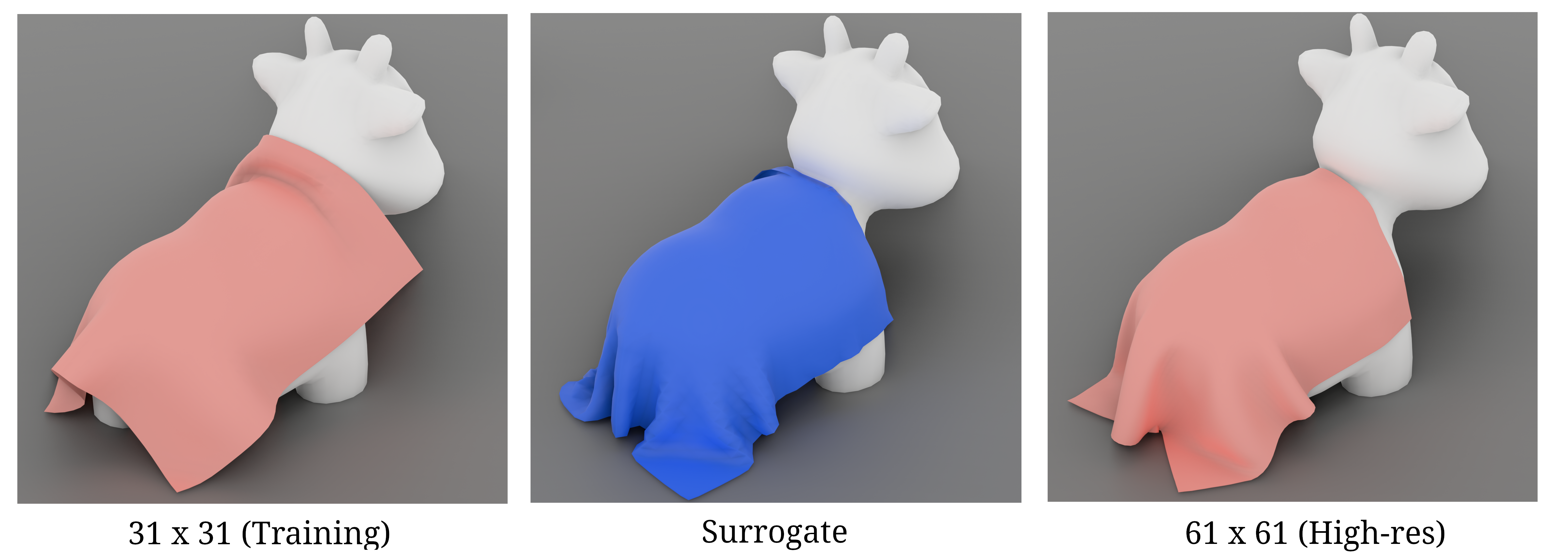}
    \caption{Generalization to novel dynamics. Left: after simulation starts, under gravity, the source cloth of low resolution ($31 \times 31$) drapes onto a cow~\citet{crane2013robust} beneath it and eventually settles to a steady state. Middle: the surrogate cloth of comparable resolution ($32 \times 32$) drapes onto a replica of the same cow. Right: the high-resolution ($61 \times 61$) source cloth drapes onto the cow for reference.}
    \label{fig:gen_novel_dyn}
\end{figure}

\textbf{Generalization to novel dynamics.} We place the trained surrogate cloth models from the above experiment under a new scene, in which all particles on the surrogate cloth are free to move, and the surrogate cloth starts falling from time $t=0 s$ until it drops onto another object. We also instantiate the source cloth in the same scene and set it to have the same initial state (position and orientation of the cloth) as the surrogate, and place an identical object below it for the source cloth to drape onto. Figs.~\ref{fig:train_vs_high_res} and \ref{fig:gen_novel_dyn} show the steady state after each system settles. We use Warp~\citep{warp2022}'s built-in SDF-based collision detector.

SpringTime produces a satisfactory match to the static, draped position of the source cloth, with fine wrinkles that can be difficult to observe if the FEM-simulated source cloth has low resolution. The Neural ODE model fails to yield a stable result during generalization, and methods like MGN cannot be easily applied to scenarios with previously unseen objects or contacts in the scene, since such methods require an input window consisting of multiple frames of kinematic data, and they do not take external forces, contact information or actions as inputs to the model. 


%% file: sections/concl.tex
We presented SpringTime, a surrogate model trained with a novel force-impulse loss and training curriculum that can learn complex, spatially-varying constitutive properties of a piece of simulated cloth without relying on a reference mesh. We demonstrated that our method can reproduce the observed behavior of both homogeneous and heterogeneous cloth sheets better than previous neural network approaches. It generalizes more readily to new scenarios, including those with previously unobserved contacts. Further, we showed that SpringTime's mass-spring-based architecture is fairly robust against membrane locking which plagues FEM-based simulations.

Our method has some limitations. While mass-spring lattices can represent a wide-range of material properties, it is uncertain if they provide universal approximating power for constitutive behavior. The topology itself could be learned for better performance and better resolution of the spatial variation. Finally, to learn spatially varying material properties, all parts of the cloth should be ```sufficiently excited'' in the training data. Better methods to achieve this than dropping the cloth in various ways could be designed, to improve the recovery of material properties.

Despite these limitations, we were pleasantly surprised by how well mass-spring systems work when trained with FEM simulation data, and their surprising ability to avoid membrane locking artifacts even when they are present in some of the training examples. Thus, when trained by our offline process, SpringTime could provide effective bridge between sophisticated and slow FEM simulators and the needs of real-time applications in games and VR.

%% file: sections/app_syn_data_gen.tex
\label{sec:app_syn_data_gen}

Other than particle positions and velocities over time, for each rollout we register
\begin{inparaenum}
    \item the particles fixed in space which is the boundary condition $\Omega$ discussed in Sec.~\ref{sec:exp_syn_data_gen};
    \item the mass of the cloth. For a simulated, dynamic source cloth this can be accessed by querying the area density $\rho$ and surface area $A$ of the cloth from the simulator used to generate data; for real-world objects we expect users to have the means of measuring the object's weight and from there they can obtain the mass of the cloth.
    \item physical dimension of the cloth, i.e. length and width. We assume the cloth always take a rectangular shape.
\end{inparaenum}

Although in Sec.~\ref{sec:method} we mentioned that external force injected into the source cloth is required as an input to our training curriculum, in practice we do not need to track it on a per-time step basis: gravity can be easily simulated for the surrogate knowing that $-9.81~\mathrm{m/s^2}$, the Rayleigh damping constant is know prior to simulation starts since this is a user-specified parameter. If we are to extend our method to source cloth that exist in the real world, we can simply ignore the damping force, e.g. air drag and assume gravity being the only external force.

\textbf{Construction of the point cloud clip dataset.} Following the approach adopted by prior work on system identification with differentiable simulators~\citep{Ma2023, gartner2022differentiable} in training, we partition each rollout into non-overlapping ``clips'' with each clip containing $T$ steps of simulation data, and make the surrogate system track the source system's motion throughout a clip on each training iteration. For training mass-spring net, we set $T=500$, so each rollout gives $\frac{8k}{500} = 16$ clips, where $8 \mathrm{s}$ of rollout gives $8k$ steps of data at timestep size of $1 \mathrm{ms}$. So from $512$ rollouts we get $8k$ clips, or $4M$ time steps of kinematic data. To reduce memory footprint and disk space usage, we regularly downsample time steps-- taking $1$ in every $40$ steps, and that leaves us $100k$ time steps of training data. We then take clips from the last $1 s$ of each rollout to yield $50k$ time steps of low-velocity training motion data.

%% file: sections/app_impl.tex
\subsection{Implementation Details}
As in ~\citet{Ma2023} we implement our neural constitutive model in PyTorch~\citep{paszke2019pytorch}, and we implement contact resolution and forward dynamics in WARP~\citep{Macklin2022}. Both forward and backward propagation through our surrogate model are fast, since CUDA tensors from PyTorch can be directly used by WARP, and we have exploited WARP's parallelism by allowing simulation of multiple clips to happen in parallel. All experiments are run on NVIDIA V100's with $32$GB of GPU memory.

For all experiments we train the neural constitutive model with the Adam optimizer. Learning rates, number of training iterations and the weight of different loss terms are task-specific.


\subsection{Hyperparameters}
\label{sec:app_hparams}
\textbf{Impulse loss vs. force loss}. When training SpringTime on kinematic data sampled from mass-spring based cloth, we used a combination of force loss and impulse loss; specifically, we set $\lambda_{\text{f}}=1$ and $\lambda_{\text{J}}=10$ in both the stiffness and the damping pass. On the other hand, for training on data sampled from mesh-based cloths, in the stiffness pass we used the impulse loss exclusively with $\lambda_{\text{J}}=1$, and in the damping pass we used the force loss exclusively with $\lambda_{\text{f}}=1$. We believe the optimal combination of the force and impulse term depends on
\begin{inparaenum}[(i)]
    \item the source of training data and
    \item whether we are estimating springs' elastic stiffness or damping properties.
\end{inparaenum}
What is particularly interesting is that for FEM-mesh-based training data, using impulse loss is crucial for learning stiffness, yet both impulse and force loss are crucial for obtaining good damping estimates. We hypothesize that impulse loss which tracks error in force accumulation over time is more resistant to noises induced by taking the second derivative with respect to particle positions, therefore more suitable for learning from low-velocity data in the stiffness pass. On the other hand, force loss is sensitive to minor perturbations induced by taking derivatives. While this is the case for learning from meshes that bear a different resolution from the surrogate, we see that in the case of learning from mass-spring sheet with the same resolution as the surrogate, both impulse and force loss enable effective learning in either the stiffness or the damping pass.

\textbf{Hyperparameter choices for the reconstruction accuracy experiment.} We train SpringTime for $192$ iterations, including $128$ iterations in the first (stiffness) pass and $64$ in the second (damping) pass. We treat each $1000$-step-long motion clip as a sample, and we simulate with batch size of $32$ clips in forward and backpropagation. Given that the training set contains $512$ clips, training for $256$ iterations in the stiffness pass is equivalent to training for $16$ epochs. For the loss function, we use $\lambda_{f}=1$, $\lambda_{J}=10$. We notice that including the force loss is critical for narrowing the distribution of stiffness and damping estimates near ground-truth values.

\textbf{Hyperparameter choices for the membrane locking experiment.} To model actual bending and show that our bending springs can model the proper bending in a source FEM mesh, we included bending springs in the surrogate. Also, we included both diagonal springs to eliminate the possibility of introducing pseudo-stiffness in the misaligned case due to asymmetry. We trained MGN~\citep{pfaff2020learning} for $3.8\mathrm{M}$ steps over 48 hours. The number of training iterations we used is much lower than the $10\mathrm{M}$ iterations~\citet{pfaff2020learning} used to train MGN on the \texttt{flag\_simple} example reported in their paper. We trained for fewer steps because 
\begin{inparaenum}[(i)]
    \item the low-resolution mesh is more coarse than the \texttt{flag\_simple} example, as our mesh contains roughly only $0.5\times$ vertices as the mesh they used;
    \item time is the constraining factor-- training for $10\mathrm{M}$ steps would require more than $126$ hours on a server with NVIDIA V100 which is prohibitively long for our purpose.
\end{inparaenum}
The excessively long training time can be partially attributed to lacking of parallelism in the training curriculum, e.g. simulations can only be run in batches of $1$ rollout; but also can be attributed to not leveraging sufficient physical priors for inferring constitutive properties. Those being said, for a qualitative assessment to demonstrate that MGN can be vulnerable to membrane locking in training data, the results from training after $830\mathrm{k}$ iterations should be sufficient, because we expect that after training for more iterations the results will be more visually close to the low-resolution source system's equilibrium configuration, and we do not see evidence of MGN learning to resist membrane locking throughout the course of training.

\textbf{Hyperparameter choices for the generalization experiment in which we subject SpringTime and the baselines under novel initial and boundary conditions.} The Neural ODE baseline contains $6$ hidden layers, with $512$ neurons in each hidden layer, and we use the exponential ReLU activation after each hidden layer. We train SpringTime for $256$ iterations in the stiffness pass, and for $64$ iterations in the damping pass, until loss converges. we train Neural ODE for $256$ iterations, with the same loss function as used for SpringTime, on a single pass, and we have made sure that training loss had converged. We train MGN for $3.8$ million iterations. But note that training of MGN works a bit differently:
\begin{inparaenum}[(i)]
    \item we train on $512$ rollouts with $8000$-step long clips. The downsampling strategy introduced in Sec.~\ref{sec:app_syn_data_gen} is applied here as well, so we get $200$ frames of data per clip. This is on the same scale as $250$ frames of data per clip used in training the \texttt{flag\_dynamic} example in MGN~\citet{pfaff2020learning}.
    \item training uses batch size of $1$ clip.
\end{inparaenum}
Turning to the loss function: we use the impulse loss exclusively, and we found that force loss is not particularly useful in this case. We believe this is because when the source cloth is constructed from triangular meshes, elastic and damping forces that act on each particle comes from edges which has no one-to-one correspondence to springs in the surrogate. Therefore learning per-spring elastic and damping force from the net force applied onto each particle becomes difficult. Impulse loss on the other hand is more robust to the change in geometric construction.

%% file: sections/app_add_results.tex
\subsection{Generalizability to Novel Initial and Boundary Conditions}
Here we showcase more results from the experiment in which we assess the generalizability of different neural models to novel initial and boundary conditions.

\begin{figure*}[ht]
    \centering
    \includegraphics[width=.9\textwidth]{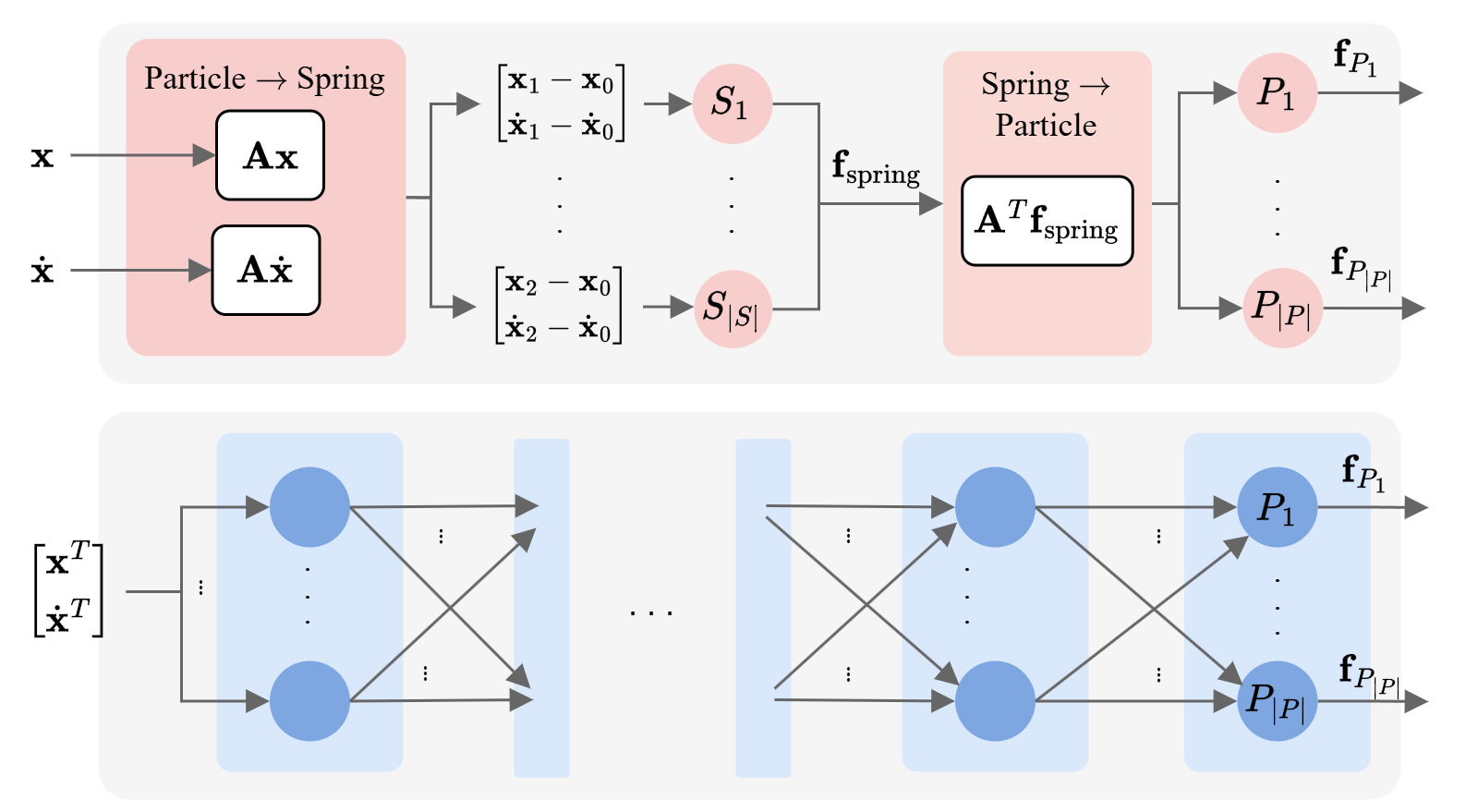}
    \caption{We illustrate the architecture of (bottom) Neural ODE, which is an MLP-based network and compare it with (top) SpringTime. Neural ODE has no springs, and neurons in the last layer directly predict constitutive forces that act on each particle in the surrogate system at any time step. Some hidden layers and connections are omitted for brevity.}
    \label{fig:arch_ours_vs_mlp}
\end{figure*}
Fig.~\ref{fig:arch_ours_vs_mlp} shows the architecture of the $512 \times 6$ MLP-based Neural ODE that we compare with SpringTime as a baseline. Without topological structure of the sheet explicitly injected into the architecture as a prior knowledge, the neural network needs to learn it by itself.

\begin{figure*}[ht]
    \centering
    \includegraphics[width=.9\textwidth]{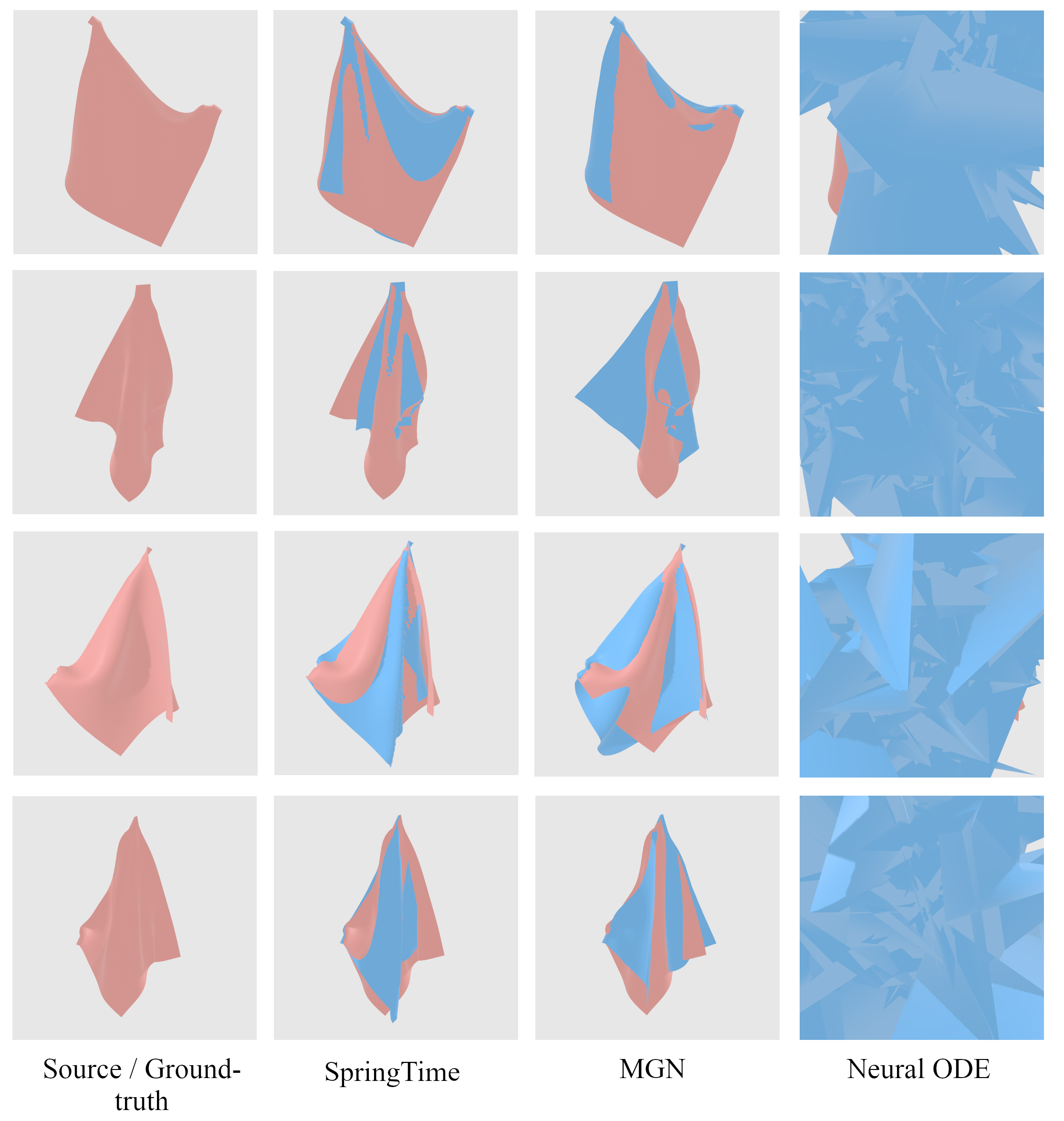}
    \caption{Generalization of SpringTime to novel initial and boundary conditions: steady states reached by the source cloth, SpringTime, MGN and Neural ODE-based surrogate after $8$ seconds, in four different test rollouts. Ground-truth cloth is in red; surrogates are in blue.}
    \label{fig:gen_novel_ic_bc}
\end{figure*}

Fig.~\ref{fig:gen_novel_ic_bc} shows the source and surrogate sheets reaching four different steady states, after four $8$-seconds-long rollouts started from four different initial and boundary conditions. Note that for each rollout, all surrogates start from the same condition as the source. Notice that our method yields significantly lower steady state errors than Neural ODE in particular, for which the prediction completely disintegrates after $8$ seconds; MGN performs comparably to SpringTime but requires considerably training time to obtain the shown results. Nevertheless, our method still has room for improvement. Notice that in the first row in Fig.~\ref{fig:gen_novel_ic_bc}, the bottom-left corner of the cloth is poorly fitted. Also notice that in the third row the mass-spring based cloth is further stretched downward than the ground-truth.


\subsection{Ablation Studies}
\label{sec:exp_ablation}
\textbf{Dual-pass vs single-pass training}. An obvious alternative to the dual-pass training curriculum would be to learn stiffness and damping constants at once, in a single pass. We conducted a single-pass training session of $128$ iterations, with batch size of $32$ clips on mass-spring cloth data, and we found that compared with results shown in Fig.~\ref{fig:recon_acc} the RMSE of stiffness constant estimates increased from $14.3$ to $22.8$, and the RMSE of damping constant estimates increased from $1.9$ to $8.5$. This is expected because of the entanglement between elastic and damping force as discussed in Sec.~\ref{sec:train_curriculum}.

\textbf{Force-and-impulse loss vs position loss.} Position-based loss which directly penalizes step-wise Euclidean distance between target and predicted particle positions is commonly used in neural constitutive modelling~\citep{Bianchi2003, Ma2023, zong2023neural}. We tried learning stiffness with position loss $|| \mathbf{x} - \hat{\mathbf{x}} ||^2$ on mass-spring cloth data. We use regularization strength $\lambda_{\text{k neg.}} = 100$, since position loss tends to be larger than force or impulse loss. Other hyperparameters are the same as when using force-and-impulse loss. We notice that with position loss, the RMSE of stiffness estimates increased significantly from $14.3$ to $26.3$. We believe what makes force-and-impulse loss or specifically, force loss working really well when the source cloth is made up of masses and springs is that it is highly sensitive to stepwise difference between predicted and target elastic and damping forces. Position loss like impulse loss effectively integrates errors over time, and the accumulation of per-step errors can make the neural model find it difficult to estimate system parameters on a finer scale. 